\documentclass[twocolumn,prl,showpacs]{revtex4}
\usepackage{epsf}
\usepackage[usenames]{color}
\newcommand{\comment}[1]{}
\usepackage{graphicx}
\usepackage{amsmath}
\usepackage{epsfig}
\usepackage{eufrak}

\newcommand{\comm}[2]{\left[ {#1},{#2} \right]}
\newcommand{\cn}[2]{c_{{#1},{#2}}}
\newcommand{\cy}[2]{c^\dagger_{{#1},{#2}}}
\newcommand{\dn}[2]{c_{{#1}+{\bf Q},{#2}}}
\newcommand{\dy}[2]{c^\dagger_{{#1}+{\bf Q},{#2}}}

\def\be{\begin{eqnarray}}
\def\ee{\end{eqnarray}}
\def\nn{\nonumber}
\def\l{\left}
\def\rr{\right}

\def\bk{{\bf k}}
\def\half{\frac{1}{2}}
\def\up{\uparrow}
\def\down{\downarrow}

\begin{document}
\title{Role of electronic nematicity in the interplay between s- and  d-wave broken-symmetry states}
%\title{Complexity induced by nematicity; d-waveness induced by electronic nematic order}
\author{Hae-Young Kee}
%\email{hykee@physics.utoronto.ca}
\affiliation{Department of Physics, University of Toronto, Toronto,
Ontario M5S 1A7 Canada}
\affiliation{Korea Institute for Advanced Study, Seoul, Korea }
\date{\today}
\begin{abstract}
To understand the role of electronic nematic order in the interplay between  s- and d-wave particle-particle or 
particle-hole condensate states,  relations between  various s- and d-wave order parameters  are studied.
We find that the nematic operator transforms two independent six-dimensional vectors.
The d-wave superconducting,  d-density wave, and antiferromagnetic orders are organized into one vector,
and the s-wave superconducting, charge density wave, and spin-triplet d-density wave orders into
the other vector.  Each vector acts as a superspin and transforms under the action of SO(6)  where
charge, spin, $\eta$- and $\pi$-pairing, spin-triplet nematic operators satisfy the SO(6) Lie algebra.
Electronic nematic order is not a part of the SO(6) group. It commutes with all 15 generators.
Our findings imply that  nematic order does not affect the competition among the order parameters
within the same superspin, while it strongly interferes the interplay between two order parameters that belong to different
superspins. For example, nematicity allows a linear coupling between d- and s-wave superconducting
order parameters which modifies the superconducting transition temperature.
A generalized Ginzburg-Landau theory and further physical implications  are discussed.

\end{abstract}
\maketitle

\section{Introduction} 
A minimal model for strongly correlated materials including the most complex systems such
as the high temperature cuprates is the Hubbard or t-J model.  However, even within the simplified Hubbard model
away from the half-filling, there has been no consensus about  the ground states, and various theoretical proposals
have been made for the phase diagram of the high temperature cuprates.

One of such states, the staggered flux phase (d-density wave), was discussed in 
exitonic condensation\cite{halperinSSP68}, t-J model, and Hubbard model\cite{kotliarPRB88,marstonPRB89,
hsuPRB91,lee-review07},
and further suggested as a pseudogap phase of the cuprates\cite{chakravartyPRB02}.
The spin-triplet version of the staggered flux phase (spin-triplet d-density wave)  was also discussed in the context of high temperature cuprates.\cite{doraPRB07,virosztekIJMP02,keeEPL09}.

Another proposed particle-hole condensate state of the angular momentum $l$=2 channel 
with broken rotational symmetry  is the electronic nematic state.\cite{kivelsonNATURE98,yamaseJPSJ00,metznerPRL00,oganesyanPRB01,keePRB03,khavkinePRB04,kivelsonRMP03,fradkin09104166,vojtaAP09}.
The spontaneous formation of nematicity has been invoked to explain the anisotropic transport
observed in a two-dimensional electron gas in a high
magnetic field\cite{lillyPRL99,duSSC99} and in Ru-based oxides\cite{borziSCIENCE07}.
Its relevance to high temperature cuprates was evidenced by a neutron scattering measurement
where anisotropic scattering patterns have been observed in YBa$_2$Cu$_3$O$_{6.5}$.\cite{hinkovSCIENCE08} 
From a weak-coupling point of view, it is  sometimes also called Pomeranchuk instability\cite{pomeranchuk}.

Given various proposed order parameters, the interplay between s- and d-wave order parameters  has been of intensive theoretical study.
%since the first observation of the exact symmetry of the Hubbard model made by Yang\cite{yangPRL89}.
Examples for s- and d-wave order parameters  are s- and d-wave superconductors, charge density wave,
spin density wave, d-density wave,  spin-triplet d-density wave, and nematic  states.   
Among them,  it was reported that nematicity  plays an important role in the interplay between
 s- and d-wave superconductors\cite{keeJPCM08}, and the spin density wave and spin-triplet  d-density wave states
\cite{keeEPL09}.  However, a full set of the relations between them is still lacking. 

In this paper, we offer a complete theory on how
s- and d-wave orders transform via the nematic order, and relations between them.
We found that the order parameters listed above can be organized in two independent six-dimensional vectors.
One vector is composed
of d-wave superconducting, d-density wave, and spin density wave order parameters, while the other vector
contains s-wave superconducting, charge density wave, and spin-triplet d-density wave order parameters.
Each vector transforms under the action of SO(6).  
Charge, spin, $\eta$-pairing
\cite{yangPRL89}, $\pi$-pairing\cite{zhangSCIENCE97}, and spin-triplet nematic\cite{wuPRB07} operators
together satisfy the SO(6) Lie algebra.

The nematic order parameter  commutes with all generators, and hence is not a part of the SO(6) group.
However, it transforms the two independent vectors connecting s- and d-wave order parameters. 
Our findings imply that nematic order does not interfere the competition between the order parameters within
the same vector, but strongly affects the interplay between two order parameters belonging  to different vectors. For example, it allows a linear coupling between s- and d-wave order parameters
in the Ginzburg-Landau (GL) free energy,  which modifies the physical properties of both phases.
We found that the  conditions for  non-zero linear coupling differ for particle-particle and particle-hole condensate
states.

Below we will review an SO(6) group theory and 
present the  relations between the nematic order,  the generators, and superspins. 
We will show the role of nematic order for particle-particle condensate states, 
and elaborate a similar process for particle-hole cases. 
 We will also discuss the implications of such relations
in the context of GL free energy, and superconducting transition temperature 
related to high temperature cuprates.

%In particular, d-wave superconductors found to be tuned from antiferromagnetic states by applying
%pressure or introducing holes or chemical substition.\cite{heavy-fermion,organic,cuprates}

\section{ nematic order parameter and SO(6) group}  
It was first pointed out by Yang\cite{yangPRL89} that the $\eta$-pairing state is an eigenstate of the Hubbard model. 
The $\eta$-pairing operator is defined as
$\hat{\eta}^+ =-i\sum_{\bk\sigma\sigma'} \cy{\bk}{\sigma}\sigma^y_{\sigma\sigma'}\dy{-\bk}{\sigma'}$ and $\hat{\eta}^-=(\hat{\eta}^+)^\dagger$, where ${\bf Q} =(\pi,\pi)$ and $\sigma^y$ is a Pauli matrix. 
The $\eta$ operator carries charge $2e$ and spin 0,
 and commutes with the Hubbard Hamiltonian
 at half filling $\mu = U/2$,  where $\mu$ is the chemical potential and $U$ is the on-site Hubbard interaction.
 It is also an eigenstate of the momentum operator with the eigenvalue ${\bf Q}$.
 
Later  it was  found that the $\eta$-pairing operators combined with the charge  operator satisfy  an SU(2) algebra (named pseudospin)\cite{zhangPRL90}, 
and further recognized that the Hubbard model has two sets of commuting SU(2) symmetries. One set is characterized by
the pseudospin of $\eta$-pairing and charge operators, and the other is conventional spin operator.\cite{yang-zhang}
The three-dimensional vector transforming under the action of pseudospin SU(2)  forms a superspin, 
where its three components are
s-wave superconductor, and charge density wave with the ordering wave vector ${\bf Q}$. 
It was also reported that the pseudospin SU(2) rotates another superspin 
composed of d-wave superconducting and d-density wave order parameters.\cite{nayakPRB00}

On the other hand,
the vector transforming under  the spin SU(2) is the spin density wave with the ordering
wave vector ${\bf Q}$.  Under a particle-hole transformation for one spin species, $c^\dagger_{i \downarrow}  \rightarrow   (-1)^i c_{i \downarrow}$, the role of the
two sets of SU(2) generators is interchanged.  The same particle-hole transformation maps the positive Hubbard model
to the negative Hubbard model, and it also maps the $\eta$-pairing to the Nagaoka ferromagnetic state.\cite{singhPRL91}
A further generalization of the concept of the exact SO(4) symmetry of the Hubbard model
to a unified theory of antiferromagnetism and d-wave superconductivity based on
SO(5) symmetry  was later proposed to understand the physics of the high temperature cuprates.\cite{zhangSCIENCE97,demlerRMP04}

%Similarly the spin SU(2) generators rotate a superspin representing the spin-triplet staggered flux state (
%triplet version of d-density wave state) defined as
%$\hat{\Delta}_{tsf}^\alpha= \frac{i}{2} \sum_{\bk\sigma\sigma'} d({\bf k}) \l(\cy{\bk}{\sigma}\sigma^\alpha_{\sigma\sigma'}\dn{\bk}{\sigma'}-\dy{\bk}{\sigma}\sigma^\alpha_{\sigma\sigma'}\cn{\bk}{\sigma}\rr) $,
%where $d({\bf k}) = \cos{k_x}-\cos{k_y}$.

Here we present a full list of relations between s- and d-wave order parameters including those mentioned above.
%-- s- and d-wave superconductors, d-density wave, charge density wave, spin density wave, and spin-triplet d-density wave states.
The ground state order parameters transform as a six-dimensional vector under the action of SO(6).
They  can be organized into a vector  ${\hat n}_a$  ($a = 1...6$) which should satisfy
\begin{equation}
[{\hat L}_{ab}, {\hat n}_c] = -i \left( \delta_{bc} {\hat n}_a -\delta_{ac} {\hat n}_b \right).
\end{equation}
where ${\hat L}_{ab}$ are generators of SO(6) based on the following
operators:
\begin{eqnarray}
\label{eq:generators}
\hat{Q} &=&-\half\sum_{\bk\sigma}\l( \cy{\bk}{\sigma}\cn{\bk}{\sigma} +\dy{\bk}{\sigma}\dn{\bk}{\sigma}-1\rr), \cr
\hat{S}_\alpha &=&\half\sum_{\bk\sigma\sigma'}\l( \cy{\bk}{\sigma}\sigma^\alpha_{\sigma\sigma'}\cn{\bk}{\sigma'} +\dy{\bk}{\sigma}\sigma^\alpha_{\sigma\sigma'}\dn{\bk}{\sigma'}\rr), \cr
\hat{R}_\alpha &=&\half\sum_{\bk\sigma\sigma'} d({\bf k}) \l( \cy{\bk}{\sigma}\sigma^\alpha_{\sigma\sigma'}\cn{\bk}{\sigma'} -\dy{\bk}{\sigma}\sigma^\alpha_{\sigma\sigma'}\dn{\bk}{\sigma'}\rr), \cr
\hat{\Pi}^+{_\alpha} &=&\sum_{\bk\sigma\sigma'}  d({\bf k}) \cy{\bk}{\sigma}\l(\sigma^\alpha  \sigma^y\rr)_{\sigma\sigma'}\dy{-\bk}{\sigma'},\ \ \ \hat{\Pi}^-=(\hat{\Pi}^+)^\dagger, \cr
%\hat{\zeta}_\alpha &&=\sum_{\bk\sigma\sigma'} \dn{-\bk}{\sigma}\l(\hat{\sigma}_y\hat{\sigma}_\alpha\rr)_{\sigma\sigma'}\cn{\bk}{\sigma'} \cr
\hat{\eta}^+ &=&-i\sum_{\bk\sigma\sigma'} \cy{\bk}{\sigma}\sigma^y_{\sigma\sigma'}\dy{-\bk}{\sigma'},\ \ \ \ \ \ \hat{\eta}^-=(\hat{\eta}^+)^\dagger,
%\hat{\eta} &&= i\sum_{\bk\sigma\sigma'} \dn{-\bk}{\sigma}\sigma^y_{\sigma\sigma'}\cn{\bk}{\sigma'}
\label{generators}
\end{eqnarray}
where ${\bf k}$ runs over the reduced Brillouin zone, $d({\bf k}) = \cos{k_x} -\cos{k_y}$,
$\alpha$ takes the values $x$, $y$, $z$, and
${\sigma}^{\alpha}$ are the Pauli matrices.
%Note that $\hat{\Pi}$ operators represent $\pi$-pairing\cite{zhangSCIENCE97}, $\eta$ operators $\eta$-pairing\cite{yangPRL89},
%and $\hat{N_t}$ operators spin-triplet version of nematic order\cite{wuPRB07}.

The generators of SO(6) can be represented by an antisymmetric 6x6 matrix, 
$\hat{L}_{ab}=-\hat{L}_{ba}$. 
\begin{eqnarray}
\hat{L}_{ab}=\left(\begin{array}{c c c c c c}
0 & \hat{Q} & \Re\,\hat{\Pi}_x & \Re\,\hat{\Pi}_y & \Re\,\hat{\Pi}_z & \Re\,\hat{\eta} \\
\, & 0 & \Im\,\hat{\Pi}_x & \Im\,\hat{\Pi}_y & \Im\,\hat{\Pi}_z & \Im\,\hat{\eta} \\
\, & \, & 0 & \hat{S}_z & -\hat{S}_y & \hat{R}_x \\
\, & \, & \, & 0 & \hat{S}_x &  \hat{R}_y \\
\, & \, & \, & \, & 0 & \hat{R}_z \\
\, & \, & \, & \, & \, & 0
\end{array}
\right),
\label{eq:genMatrix}
\end{eqnarray}
where $\Re\,\hat{\cal O}\equiv\half(\hat{\cal O}^-+\hat{\cal O}^+)$ and $\Im\,\hat{\cal O}\equiv\frac{1}{2i}(\hat{\cal O}^--\hat{\cal O}^+)$. 
It satisfies the correct SO(6) Lie algebra,\cite{footnote}
\begin{eqnarray}
\label{eq:LLComm}
\comm{\hat{L}_{ab}}{\hat{L}_{cd}}=-i\l(\delta_{ad}\hat{L}_{bc}+\delta_{bc}\hat{L}_{ad}-\delta_{bd}\hat{L}_{ac}-\delta_{ac}\hat{L}_{bd}\rr).\nn
\label{so-6-group}
\end{eqnarray}
Here $L_{12}$ is the charge operator, and $L_{34}$,$L_{35}$, and $L_{45}$ are the  three components of the spin operator.
$L_{16}$ and $L_{26}$ represent real and imaginary part of the $\eta$-pairing.\cite{yangPRL89}
% $\eta = c_{{\bf k},\uparrow} c_{-{\bf k}+{\bf Q},
%\downarrow} $ where ${\bf Q}+(\pi,pi)$, and  $(\eta)^+ = (\eta)^\dagger$.
%The $\eta$-pairing operator carries charge $2e$ and spin 0, and represents a broken translational symmetry.
$L_{13}$, $L_{14}$, and $L_{15}$ ($L_{23}$, $L_{24}$ and $L_{25}$) denote $x$, $y$ and $z$ component of the real (imaginary) part of 
the $\pi$-pairing which carries charge $2e$ and spin $1$ and represents a broken translational symmetry.\cite{zhangSCIENCE97}
%$\pi_{i} = d({\bf k}) c_{{\bf k},\alpha} \sigma^{i}_{\alpha,\beta} c_{-{\bf k}+{\bf Q}, \beta}$
%where $i=x,y,z$ and $\sigma$ is Pauli matrix, and $(\pi^{i})^+ = (\pi^{i})^\dagger$\cite{sczhang}
$L_{36}$, $L_{46}$, and $L_{56}$ correspond to the spin-triplet nematic order parameter carrying spin 1
and representing a  broken x-y symmetry on a square lattice.\cite{wuPRB07}

There exist two independent vectors. Each vector acts as a superspin,
and transforms under the SO(6). This observation was first reported in Ref. \cite{markiewicz97},
and a similar SO(6) symmetry was found in Fe-pnictide superconductors\cite{podolskyEPL09}.
One superspin (superspin-1) consists of  spin-density wave ($\Delta_{sdw}$),  d-density wave ($\Delta_{ddw}$), and  d-wave superconducting ($\Delta_{dsc}$)
order parameters:
\begin{eqnarray}
\hat{\Delta}_{sdw}^\alpha&=&\half\sum_{\bk\sigma\sigma'}\l(\cy{\bk}{\sigma}\sigma^\alpha_{\sigma\sigma'}\dn{\bk}{\sigma'}+\dy{\bk}{\sigma}\sigma^\alpha_{\sigma\sigma'}\cn{\bk}{\sigma}\rr), \nonumber\\
\hat{\Delta}_{dsc}^+ &= &\sum_{\bf k} d({\bf k}) \l(\cy{\bk}{\up}\cy{-\bk}{\down}-\dy{\bk}{\up}\dy{-\bk}{\down}\rr),\nonumber\\\hat{\Delta}_{dsc}^-&=& (\hat{\Delta}_{dsc}^+)^\dagger,\nonumber \\
\hat{\Delta}_{ddw}&=&-\frac{i}{2}\sum_{\bk\sigma} d({\bf k}) \l(\cy{\bk}{\sigma}\dn{\bk}{\sigma}-\dy{\bk}{\sigma}\cn{\bk}{\sigma}\rr),
\end{eqnarray}
where ${\hat n}_1 = \Re\, \hat{\Delta}_{dsc}$, ${\hat n}_2 = \Im\, \hat{\Delta}_{dsc}$, $ {\hat n}_3 = \hat{\Delta}_{sdw}^x$, $ {\hat n}_4
=\hat{\Delta}_{sdw}^y$, ${\hat n}_5 = \hat{\Delta}_{sdw}^z$, and ${\hat n}_6 = \hat{\Delta}_{ddw}$.

The other superspin (superspin-2) rotated by the same 15 generators ${\hat L}_{ab}$ 
is composed of spin-triplet d-density wave ($\Delta_{tsf}$),  charge density wave ($\Delta_{cdw}$), and s-wave superconducting ($\Delta_{ssc}$) order
parameters:
\begin{eqnarray}
\hat{\Delta}_{tsf}^\alpha&=& \frac{i}{2} \sum_{\bk\sigma\sigma'} d({\bf k}) \l(\cy{\bk}{\sigma}\sigma^\alpha_{\sigma\sigma'}\dn{\bk}{\sigma'}-\dy{\bk}{\sigma}\sigma^\alpha_{\sigma\sigma'}\cn{\bk}{\sigma}\rr), \nonumber\\
\hat{\Delta}_{ssc}^+ &= &\sum_{\bf k} \l(\cy{\bk}{\up}\cy{-\bk}{\down}+\dy{\bk}{\up}\dy{-\bk}{\down}\rr), \nonumber\\
 \hat{\Delta}_{ssc}^-&=&(\hat{\Delta}_{ssc}^+)^\dagger,\nonumber \\
\hat{\Delta}_{cdw}&=&-\frac{1}{2}\sum_{\bk\sigma} \l(\cy{\bk}{\sigma}\dn{\bk}{\sigma}+\dy{\bk}{\sigma}\cn{\bk}{\sigma}\rr),
\end{eqnarray}
where the superspin is arranged as
${\hat n}_1 = \Re\, \hat{\Delta}_{ssc}$, ${\hat n}_2 = \Im\, \hat{\Delta}_{ssc}$, $ {\hat n}_3 = \hat{\Delta}_{tsf}^x$, $ {\hat n}_4
=\hat{\Delta}_{tsf}^y$, ${\hat n}_5 = \hat{\Delta}_{tsf}^z$, and ${\hat n}_6 = \hat{\Delta}_{cdw}$.

What is the role of  the nematic order parameter  within the SO(6) group?
The nematic order parameter is given by
\begin{equation}
{\hat N} = \sum_{{\bf k}\sigma} d({\bf k}) \l( \cy{\bk}{\sigma} \cn{\bk}{\sigma} - \dy{\bk}{\sigma} \dn{\bk}{\sigma}  \rr).
\end{equation}
When $\langle {\hat N} \rangle \equiv {\cal N}_0 \neq 0$, the phase is characterized by a broken x-y symmetry of the square lattice, and it is trivial to  generalize to a broken point group symmetry in other lattices.
%Let us study relations between the nematic order and other broken symmetric states in $l=0$ and $l=2$ channels.
%Below we will first show a relation between the nematic order and the generators which form
%an SO(6) transformation group\cite{Markiewticz} which includs a subset of SO(5) proposed for 
%high temperature superconductivity in the cuprates\cite{sczhang}
%
Note that the nematic order parameter commutes with all 15 generators:
\begin{equation}
[\hat{N}, {\hat L}_{ab}]=0.
\label{nematic-generator}
\end{equation}
This means that nematicity is not an SO(6) symmetry breaking field, and does not
interfere with the competition between the order parameters within the superspin.
For example, the phase diagram between antiferromagnetism and d-wave superconductivity
(both belong to superspin-1)
studied in the t-J model based on SO(5) symmetry\cite{zhangSCIENCE97} (a subset of the SO(6) in this study) 
is not modified by  the presence of nematicity.

However, the nematic operator does not commute with the following conventional quantum rotor model.
\begin{equation}
H_{QR} = \frac{1}{2\chi} \sum_{i, a< b} {\hat L}_{i,ab}^2 + \sum_{<i j>,a} r_a {\hat n}_i^a {\hat n}_j^a,
\end{equation}
where the first term is the kinetic term and  $\chi$ is the moment of intertia, and the second term
is the potential term. The Hamiltonian has SO(6) symmetry when $r_a$ is idential to all $a$.
 Note that the nematic operator  commutes with the first term, but not the second term.
On the other hand, the competition between the order parameters in the same superspin ${\hat n}^a$
is determined by difference in $r_a$.

%To understand the physical implication of the above result, let us introduce
%a quantum rotor model\cite{demlerRMP04} which has the SO(6) symmetry.
%\begin{equation}
%H_{QR} = \frac{1}{2 \chi} \sum_{i,a < b} {\hat L }_{i,ab}^2 - r \sum_{<ij>,a} {\hat n}_{i,a} {\hat n}_{j,a},
%\end{equation}
%Then note that adding a term such as $H_{nem} = -g \sum_{ij} {\hat N}_i {\hat N}_j$
%which favors the nematic state, does not break the SO(6) symmetry, because
%the generators ${\hat L}_{ab}$ commute with ${\hat N}$.
%Therefore, the nematic state does not interfere the interplay among the order parameters forming the superspin-1
%(or superspin-2). In other words, the mechanism of forming the nematic state is not shared by the mechanism of 
%the superspin orders ${\hat n}_{a}$, and can exist independent of them.
%However, when the nematic order exists, it leads to complexity since it allows to couple two ${\hat n}_{a}$ in superspin-1 and -2.
%
%
%On the other hand,  the nematicity leads a complex phase diagram,
%as it allows a link between two different superspins, as we will discuss below.
%
%{\it superspin, order parameters and their relation to nematic order}
%The relations between the nematic order parameter and the superspin rotated under SO(6) are rather different.
What are relations between nematic order and other order parameters?
%Is there any relation between nematic order and the two superspins?
The nematic operator transforms the components of the two independent superspins as follows:
\begin{eqnarray}
\left[ \hat{\Delta}_{dsc}^+ , \hat{\Delta}_{ssc}^- \right] &=&  2 \hat{N},
\;\;\;\;\;  \left[ \hat{\Delta}_{dsc}^-, \hat{\Delta}_{ssc}^+ \right]   =  2 \hat{N},\nonumber
\\
\left[ \hat{\Delta}_{ddw},\hat{\Delta}_{cdw} \right] &=& i \hat{N},
\;\;\;\;\; 
\left[ \hat{\Delta}_{tsf}^z,\hat{\Delta}_{sdw}^z \right] = i \hat{N}, \nonumber
\\
\left[ \hat{\Delta}_{tsf}^+,\hat{\Delta}_{sdw}^- \right] &=& \frac{i}{2}  \hat{N},
\;\;\;\;\;\; 
\left[ \hat{\Delta}_{tsf}^-,\hat{\Delta}^+_{sdw} \right] = \frac{i}{2} \hat{N}.
\label{nematic-orders}
\end{eqnarray}
The nematic operator transforms s- to d-wave order parameters  which belong to
 different superspins. 
 
 The above results are summarized in
the table below.
%
% Requires the booktabs if the memoir class is not being used
%\begin{table}[htbp]
  % \centering
   %\topcaption{Table captions are better up top} % requires the topcapt package
  % \begin{tabular}{@{} lcr @{}} % Column formatting, @{} suppresses leading/trailing space
%      \toprule
%      \multicolumn{2}{c}{Item} \\
%      %\cmidrule(l){1-2} 
  %    % Partial rule. (r) trims the line a little bit on the right; (l) & (lr) also possible
  %    Animal    & Description & Price (\$)\\
   % %  \midrule
  %    Gnat      & per gram & 13.65 \\
  %              & each     &  0.01 \\
   %   Gnu       & stuffed  & 92.50 \\
     % Emu       & stuffed  & 33.33 \\
     % Armadillo & frozen   &  8.99 \\
   %  % \bottomrule
 %  \end{tabular}
 %  \caption{Remember, \emph{never} use vertical lines in tables.}
%   \label{tab:booktabs}
%\end{table}
%
%\begin{widetext}
\begin{table}[htbp]
  \centering
 %\topcaption{A summary of SO(6) group and its relation to the nematic order}
  \begin{tabular}{@{} cccc @{}}
  \hline
    SO(6) generators & \;\; ${\hat Q}$, ${\hat S}$, ${\hat \eta}$-, ${\hat \pi}$-pairing, spin nematic operators \\
    \hline
     nematic operator &  commutes with generators \&\\
      &   transforms superspin-1 and -2 \\
     \hline
     superspin-1 &  dSC, dDW, SDW \\
     superspin-2 &  sSC,  CDW, spin-triplet dDW\\
     \hline
        %\midrule
   % \bottomrule
  \end{tabular}
  \caption{ A summary of the SO(6) group and the relations to nematic order.}
  %dSC, sSC, dDW, SDW, CDW refer to d- and s-wave supercondutors, d-density wave, and
%spin-  and charge-density wave orders, respectively}
  \label{tab:label}
\end{table}
%\end{widetext}

In the following section, we discuss the physical implications of the commutation relations using a GL free energy theory
assuming that  nematic order is present.\cite{note}

\section{Ginzburg Landau theory}
%Let us study an implication of Eq. \ref{nematic-orders}.
The commutation relations in Eq. \ref{nematic-orders},
$[A,B]= {\hat N}$, indicate that if $\langle {\hat N} \rangle \equiv {\cal N}_0$ 
is finite, a linear coupling between A and B phases, such as $\gamma \; \Phi \; \Psi$ with $\Phi =\langle A \rangle$ and
$\Psi = \langle B \rangle $,  may  be present in the GL free energy.
The GL free energy is then given by
\begin{equation}
{\cal F} = \frac{a}{2} \Psi^2 + \frac{b}{2} \Phi^2 + \gamma \;  \Psi \; \Phi + u \Psi^4 + v \Phi^4 + ....
\label{gl-free-energy}
\end{equation}
Assuming that $a > 0$, $b > 0$, and $a b > \gamma^2$ (none of the phases represented by $\Phi$ and $\Psi$
is ordered), the solutions of the two coupled equations for $\Phi$ and $\Psi$ leads to
the following dispersion of modes:\cite{footnote2}
\begin{equation}
\chi \omega^2 ({\bf k})= \frac{\epsilon_1({\bf k})+\epsilon_2({\bf k})}{2} \pm \frac{1}{2} \sqrt{ 
(\epsilon_1({\bf k})-\epsilon_2({\bf k}))^2 + 4 \gamma^2},
\end{equation}
where 
\begin{eqnarray}
\epsilon_1({\bf k}) &= & a + \rho \{(1+ {\cal N}_0) k_x^2+(1-{\cal N}_0) k_y^2\},
\nonumber\\
\epsilon_2({\bf k}) &= & b + \rho \{(1+{\cal  N}_0) k_x^2+(1-{\cal N}_0) k_y^2\}.
\end{eqnarray}
Here we have used the effective Lagrangian 
$L_{eff} = \frac{\chi}{2} (\partial_t \Phi)^2
-\frac{\rho}{2} \{(1+{\cal N}_0) (\partial_a \Phi)^2 + (1-{\cal N}_0) (\partial_y \Phi)^2 \} - \frac{a}{2} \Phi^2
 +\frac{\chi}{2} (\partial_t \Psi)^2
-\frac{\rho}{2} \{(1+{\cal N}_0) (\partial_a \Psi)^2 + (1-{\cal N}_0) (\partial_y \Psi)^2 \} - \frac{b}{2} \Psi^2
-\gamma \Psi \Phi$, where $\rho$ is the stiffness.
Note that the excitations are anisotropic due to nematicity\cite{kaoPRB05}, and $k_x$ and $k_y$ are deviations from
an ordering wave-vector  which is either 0 or ${\bf Q}$ depending on the nature of $\Psi$ (or $\Phi$).

One of the solutions becomes 0  when $\gamma= \sqrt{a b}$, leading to an ordered state.
The condensed state is a linear combination of $\Psi$ and $\Phi$, and the dominant
contribution depends on $a$ and $b$. Also, if one of them, say $\Psi$, is finite (when $a < 0$), the other, $\Phi$, is always induced as long as $\gamma$ is finite.

Is $\gamma$ always finite if nematic order exists?
For example,  consider a system in the nematic state with SO(6) symmetry at high temperatures.
At low energy, the system spontaneously breaks the SO(6) symmetry, and one of the phases represented by
$\Psi$ is stabilized. If $\Psi$ represents the d-wave superconducting state, $\Phi$ is the s-wave component.
Similarly if $\Psi$ is the spin density wave, $\Phi$ should be the spin triplet d-density wave.\cite{keeEPL09}
Does nematicity always lead to an induced order parameter of $\Phi$
without any extra condition?
%
%Such a situation was studied in Ref. \cite{kimPRB08}, where s-wave component is proportional to
%the nematic strength.  The effects of an induced order
%parameter was presented in Ref. \cite{keeEPL09} when the spin density wave becomes a ground state.
%
%To study the collective modes, we write down the equation of motion.
%\begin{equation}
%\partial_t \left(\frac{\partial L}{\partial (\partial_t S)}\right) +\nabla \cdot 
%\frac{\partial L}{\partial (\nabla S)} -\frac{\partial L}{\partial S} =0
%\end{equation}
%
%In other words, does the above commutation relation, Eq. \ref{nematic-orders} gurantee a finite linear coupling $\gamma$?
Below we show that it requires another condition for a non-zero linear coupling (in addition to the nematic order), and
that the condition for a finite $\gamma$ differs for particle-particle and
particle-hole condensates.

\section{ Difference between particle-particle and particle-hole pairs}

Let us compute  $\gamma$ for particle-particle condensate states.
To check the condition for a non-zero linear coupling coefficient $\gamma$ between d-wave and s-wave superconducting cases
($\Psi =Re\langle \Delta_{dsc} \rangle $ and $\Phi =Re \langle \Delta_{ssc} \rangle$),
we introduce $ \psi^\dagger_{{\bf k}} =
(c^{\dagger}_{{\bf k} \sigma}, c_{-{\bf k} -\sigma})$. Then the order parameter is written as
$\Delta_{ssc} = \sum_{\bf k} \psi^{\dagger}_{\bf k} \tau_1 \psi_{\bf k}$.
Inside the nematic state, the quasiparticle Green's function is written as
\begin{equation}
G^{-1}({\bf k}, i\omega_n)= -i \omega_n + \epsilon_{\bf k}  -\mu,
\end{equation}
where
\[
\epsilon_{\bf k}= -2 t (\cos{k_x}+\cos{k_y})  + 2t d({\bf k}) {\cal N}_0 -4 t^\prime \cos{k_x}\cos{k_y},
\]
%-2t'' (\cos{2 k_x}+\cos{2 k_y})$,
and $\mu$ is the chemical potential.
$t$ and $t^{\prime}$ represent the nearest neighbor and second nearest neighbor hoppings, respectively.
Assuming that d- and s-wave superconducting fluctuations couple to fermions
with interactions of $g_1$ and $g_2$, 
 the $\gamma$ coefficient becomes
%the Feyman diagram of Fig. (\label{fig:nem-induced-order}).
\begin{eqnarray}
\gamma_{dsc-ssc} & = & g_1 g_2 T \sum_{{\bf k}} d({\bf k})\sum_{i\omega_n} Tr \left( G({\bf k},i\omega_n) \tau_1 G({\bf k}, i\omega_n) \tau_1 \right)
\nonumber\\
 & = &  g_1 g_2 \sum_{{\bf k}} d({\bf k}) \frac{ n_F(\xi_k) -n_F(-\xi_k)}{2 \xi_k},
\end{eqnarray}
where $\xi_{\bf k} = \epsilon_{\bf k} -\mu$.
$\gamma$ is always finite as long as $\mu$ and/or $t^\prime$ is finite. 
In other words, when the particle-hole symmetry is broken and nematic order is present,
the linear coupling term induces d- or s-wave superconducting order as we discussed
in Eq. \ref{gl-free-energy}.
%The effects of fluctuation of the induced s-wave component 
%in the coexistence of the nematic and d-wave superconducting phase
%was studied.\cite{eakim}

However, the above result is not true for particle-hole pairs.  $\gamma$ then is zero 
independent of particle-hole symmetry. 
To examine the condition for particle-hole cases, let us introduce $ \psi^\dagger_{{\bf k} \sigma} =
(c^{\dagger}_{{\bf k} \sigma}, c^{\dagger}_{{\bf k}+{\bf Q} \sigma})$.
In this basis, the Green's function becomes
\begin{equation}
G^{-1}({\bf k}, i\omega_n)= -i \omega_n I + \tilde{\epsilon_{\bf k}} \tau_3  -\mu_{\bf k} I ,
\end{equation}
where $\tilde{\epsilon}_{\bf k} = -2 t (\cos{k_x}+\cos{k_y})  + 2t d({\bf k}) {\cal N}_0
= -\tilde{\epsilon}_{{\bf k}+{\bf Q}}$
and $\mu_{\bf k} =  4 t^\prime \cos{k_x}\cos{k_y} + \mu = \mu_{{\bf k}+{\bf Q}}$.
The $\gamma$ coefficient for example between the charge density wave and the d-density wave order is then
obtained as
\begin{eqnarray}
\gamma_{cdw-ddw} &\propto & T \sum_{{\bf k}} d({\bf k}) \sum_{i\omega_n} Tr \left( G({\bf k},i\omega_n) \tau_1 G({\bf k}, i\omega_n) \tau_2 \right)
\nonumber\\
% & & \hspace{-2cm} = T \sum_{{\bf k}} d({\bf k}) \sum_{i\omega_n}
%Tr \left( \frac{(i \omega_n +\mu_k )I + \tilde{\epsilon}_k \tau_3}{ (i \omega_n+\mu_k)^2 -(\tilde{\epsilon}_{\bf k})^2}
%\tau_1 
%\frac{(i \omega_n +\mu_k )I + \tilde{\epsilon}_k \tau_3}{ (i \omega_n+\mu_k)^2 -(\tilde{\epsilon}_{\bf k})^2}
%\tau_2 \right)
%\nonumber\\
& = &0.
\end{eqnarray}
This is similar to the coupling between $Re \Delta_{dsc}$ and $Im \Delta_{ssc}$. This linear coupling
is not allowed in the free energy due to  symmetry.

Let us consider the coupling between different directions of spin density wave and spin-triplet d-density wave.
For example, the coupling between antiferromagnetic fluctuations along the $x$-direction
and spin-triplet d-density wave fluctuations along the $y$-direction are given by
$\delta \Delta_{sdw}^{x} \propto \psi_{\bf k}^\dagger \tau_1 \psi_{\bf k}$ and
$\delta \Delta_{tsf}^{y} \propto \psi_{\bf k}^\dagger \tau_1 \psi_{\bf k}$ 
where $\psi^{\dagger}_{\bf k} = \left( c^{\dagger}_{\bf k, \up}, c^{\dagger}_{{\bf k}+{\bf Q}, \down}
\right)$.
Then the coefficient $\gamma$ is obtained as
\begin{eqnarray}
\gamma_{tsf-sdw}  & \propto & \sum_{{\bf k} i\omega_n}  d({\bf k}) {\rm Tr} 
\left( \tau_1 G({\bf k} i\omega_n) \tau_1 G({\bf k} i\omega_n) \right)
\nonumber\\
&=& \sum_{\bf k} \frac{d({\bf k})}{2 \tilde{\epsilon}_{\bf k}} \left( 
n_F(\tilde{\epsilon}_{\bf k} -\mu_{\bf k})- n_F(-\tilde{\epsilon}_{\bf k} -\mu_{\bf k}) \right)
\nonumber\\
&=& 0.
\end{eqnarray}
Note that the coupling is also 0, because both $d({\bf k})$ and $\tilde{\epsilon}_{\bf k}$
change sign under ${\bf k} \rightarrow {\bf k}+{\bf Q}$, while $\mu_{\bf k}$ does not.
The physical reason is that the spin density wave state breaks time reversal symmetry, while
the triplet staggered flux does not. This fact is reflected in the commutation relations,
where Eq. \ref{nematic-orders} has the imaginary factor $i$. 
Therefore, a linear coupling is not allowed between s- and d-wave particle-hole condensate states.

However, in the presence of a magnetic field $h$, the result alters.
Note that the particle-particle and particle-hole order parameters are related
by the particle-hole transformation, which also maps the chemical potential to the magnetic field
to be discussed in detail below.
In the presence of an external magnetic field,  $\gamma$ changes to
\begin{eqnarray}
& & \gamma_{tsf-sdw} (h \neq 0) \\
& \propto & \sum_{\bf k} \frac{d({\bf k})}{2 (\tilde{\epsilon}_{\bf k} +h) } \left( 
n_F(\tilde{\epsilon}_{\bf k} +h -\mu)- n_F(-\tilde{\epsilon}_{\bf k}-h  -\mu) \right).
\nonumber
\end{eqnarray}
$\gamma$ between the spin-triplet d-density wave and spin density wave is finite when ${\cal N}_0$ and $h$ are finite. 
The leading contribution of $h$ and ${\cal N}_0$ to $\gamma({\cal N}_0,h)$ can be written as
$\gamma({\cal N}_0,h) = \gamma_0 {\cal N}_0 h$, where $\gamma_0$ depends on the interactions
between fermions and the fluctuations of the order parameters. \cite{keeEPL09}

To understand the difference between particle-particle and particle-hole condensates, let us consider the particle-hole
transformation. The particle-hole transformation 
%\begin{equation}
%$c^{\dagger}_{i \uparrow} \rightarrow c^{\dagger}_{i \uparrow}$ and   
%$c^{\dagger}_{i \downarrow} \rightarrow (-1)^i c_{i \downarrow} $,
%\end{equation}
mapping the positive to the negative Hubbard model discussed above
maps each component of the superspins as follows:
\begin{eqnarray}
\hat{\Delta}_{sdw}^{\alpha} &\rightarrow& \left(\hat{\Delta}^{\pm}_{ssc}, \hat{\Delta}_{cdw}\right),
\nonumber\\
\hat{\Delta}_{tsf}^{\alpha}  &\rightarrow& \left( \hat{\Delta}^{\pm}_{dsc},  \hat{\Delta}_{ddw}  \right),
\end{eqnarray}
where $\alpha= x,y,z$.

In addition to the known result that  the antiferromagnetic order
transforms to the s-wave superconducting and charge density wave orders,
we found that the spin-triplet d-density wave phase transforms to the
d-wave superconducting and d-density wave orders,
while nematic order is invariant. 
Since the chemical potential maps to the magnetic field under the particle-hole transformation,
the conditions for a finite $\gamma$ between particle-particle and particle-hole condensates
are also related by the particle-hole transformation -- $\gamma_0 {\cal N} \mu 
\langle  \Re \Delta_{dsc} \rangle \langle  \Re \Delta_{ssc} \rangle$ maps to
$\gamma_0 {\cal N} h \langle \Delta_{tsf}^y \rangle \langle \Delta_{sdw}^x  \rangle$
under the particle-hole transformation.
Therefore,  the linear coupling between 
d- and s-wave superconducting order parameters requires a finite chemical potential, while
the coupling between  spin-triplet d-density wave and spin density wave requires a magnetic
field. Note that both breaks SO(6) symmetry, as the chemical potential and magnetic field appear
as $\mu L_{12}$ and $h L_{34}$ in Hamiltonian, respectively.

%The results tell us that the d-wave superconducting phase in the repulsive Hubbard model on a nearest 
%tight binding model with a finite chemical potential maps to the triplet staggered flux 
%(where a vector is pointing in x-y plane) in the attractive Hubbard model with a finite magnetic field,
%in analogy to s-wave orders.

\section{effect on superconducting transition temperature}
Let us reexamine the  GL free energy, Eq. \ref{gl-free-energy},  to see if the superconducting transition temperature 
is modified by the coupling between the d- and s-wave superconducting order parameters.
%\begin{equation}
%{\cal F} = a \Psi^2 + b\Phi^2 + \gamma \;  \Psi \; \Phi + u \Psi^4 + v \Phi^4 + ....
%\end{equation}
We consider  $\Psi = \langle \Re \Delta_{dsc} \rangle $ and $\Phi = \langle \Re \Delta_{ssc}\rangle$, and $\gamma$ is finite and proportional to the nematic strength ${\cal N}_0$, and  particle-hole symmetry is assumed to be broken.

Since the chemical potential couples to the charge operator $L_{12}$, it favors the d-wave superconducting
state over the antiferromangetic and d-density wave states.
% as shown in Ref. \cite{zhangSCIENCE97}. 
The competition between  antiferromagnetism, d-wave superconductor, 
and d-density wave is determined by SO(6) symmetry breaking terms, where
 nematicity does not affect the interplay between them.

Assuming that the superconducting state is stabilized in a finite window of phase space,
and the transition temperature is set by $T_c^0$ ( $a < 0$ below $T_c^0$ and assume $b > 0$),
we are interested in the effect of nematicity on the superconducting transition temperature.
It is straightforward to check that the effective mass term 
$a_{eff}$ is modified by $a - \frac{\gamma^2}{4b}$ after integrating out the $\Phi$ field.
Note that $a \propto (T-T_{c}^0)$ and $a_{eff} \propto (T-T_c)$ where $T_c$ is the transition
temperature modified by the coupling $\gamma$.
Since the effective mass gets smaller due to the coupling to the s-wave component, the transition
temperature $T_c$ is higher than $T_c^0$.
However, one should note that the current description is based on a classical theory, and  quantum fluctuations
beyond the present study should be taken into account to see if the result may qualitatively change.

\section{Discussion and Summary}
We have studied the role of the nematic order parameter in the interplay between s- and d-wave particle-particle
or particle-hole condensate states. These condensate states include d- and s-wave superconductors,
d-density wave, spin-triplet d-density wave, spin density wave, and charge density wave  phases.
We found that the nematic operator transforms d- to s-wave superconductors,
 spin-triplet d-density wave to (s-wave) spin-density wave, and d-density wave to (s-wave) charge-density wave
operators.  This can be summarized as a transformation between two different six-dimensional vectors.
One vector is composed of d-wave superconductor, d-density wave, and spin-density wave 
order parameters, while the other vector consists of  s-wave superconductor, charge-density wave, and spin-triplet
d-density wave order parameters.
Each vector acts as a superspin and transforms under  the action of SO(6).  There exist 15 generators, 
which correspond to charge, spin, spin-triplet nematic,  $\eta$- and $\pi$-pairing operators, which form the SO(6)
group.  

The transformation between the two superspins via nematicity implies that a linear coupling between  two order parameters that belong to two different vectors
can be present in the GL free energy. Such a linear coupling allows  induced ordering  when one of them is condensed.
However, we found that there is an additional condition for a non-zero linear coupling, which
differs for particle-particle and particle-hole condensates. For example, 
when d-wave superconductor (particle-particle condensate) and nematic order coexist, 
  s-wave superconducting order is induced, only when the particle-hole symmetry is broken. On the other hand, when spin-density wave
(particle-hole condensate)
and nematic order coexist, a similar transformation allows an induced spin-triplet d-density wave,
only when  time-reversal symmetry is broken. These results are consistent with symmetry considerations. Since
the spin-triplet d-density wave does not break time reversal symmetry, while spin-density
wave does, a linear coupling between the two order parameters is allowed
when  time reversal symmetry is broken by an external magnetic field. 

It is also interesting to notice that the nematic operator commutes with the generators.
When  the Hamiltonian contains a term $- g \sum_{ij} {\hat N}_i {\hat N}_j$ which favors
 nematic ordering, it does not act as an SO(6) symmetry breaking field.  
It means that the nematic order can exist without interfering the competition among the six different order parameters
within a superspin. It is merely a spectator. 
However, it affects the interplay between order parameters which belong to two different superspins.
Nematicity  allows a linear coupling between the two order parameters, and
affects the physical properties
of both phases.  As an example,  we showed that the d-wave superconducting transition temperature is modified by 
the coupling
to the s-wave superconducting order parameters which happens when  nematicity is present and particle-hole
symmetry is broken.

The nematic order parameter has been widely discussed in the context of strongly correlated materials.
 In particular, the phase diagram of the high temperature cuprates is complex
and its complete understanding requires further experimental and theoretical investigation. Our results indicate
that the proposed nematic phase affects  phenomena in the superconducting phase such as an 
anisotropy in the spin susceptibility and an increase in superconducting transition temperature. It also affects
antiferromagnetism via the coupling to the spin-triplet d-density wave when a magnetic field is applied.
We do not attempt to find a microscopic  Hamiltonain with SO(6) symmetry which is beyond the scope of the current study.
However, we emphasize that 
Eq. \ref{nematic-generator} and \ref{nematic-orders} are exact independent of symmetry of Hamiltonian,
and SO(6) symmetry is useful to identify the compact  relations between the nematic
and other order parameters suggested in the context of high temperature cuprates. The GL free energy analysis
hints the importance of nematicity for the phase diagram of antiferromagnetism and d-wave superconducting phase.

{\it Acknowledgement} :  I thank D. Podolsky, M. Norman, S. Kivelson, and especially E. Fradkin 
for fruitful discussions. This work has been supported by
NSERC of Canada, Canadian Institute for Advanced Research, and Canada Research Chair.


\begin{thebibliography}{100}
\bibitem{note} Orthorhomicity from a orthorhombic crystal structure has a similar effect
as that of nematicity which refers to spontaneously broken rotational symmetry due to interactions.
\bibitem{footnote2} There exist  other collective modes as a consequence of the broken symmetry such as $\pi$-
resonance mode\cite{zhangSCIENCE97,demlerRMP04} and exotic modes discussed
in \cite{podolskyEPL09}.  However, these previous studies focused on the effects of the broken SO(5) or SO(6) symmetry, while the current study focused on the effects of induced order parameter.
\end{thebibliography}
\end{document}